\documentclass[conference,10pt,a4paper]{IEEEtran}
\usepackage{times}
\usepackage{graphicx}
\usepackage{multirow}
\usepackage[none]{hyphenat}
\usepackage[flushleft]{threeparttable}
\usepackage{float}
\usepackage{subfig}
\DeclareMathAlphabet{\mathbfit}{OML}{cmm}{b}{it}
\usepackage{cite}
\usepackage[cmex10]{amsmath}
\usepackage{color}
\usepackage{soul}
\usepackage{multicol}
\usepackage{mathrsfs,amssymb}
\usepackage{dblfloatfix}
\usepackage{makecell}
\usepackage{physics}
\usepackage{siunitx}
\usepackage{cite}
\usepackage{amsmath,amssymb,amsfonts}
\usepackage{algorithmic}
\usepackage{graphicx}
\usepackage{textcomp}
\usepackage{xcolor}
\usepackage{algorithm}

\usepackage{amsmath,amssymb}
\usepackage{url}
\usepackage{hyperref}

\allowdisplaybreaks
\makeatletter

\def\@maketitle{\newpage
\bgroup\par\addvspace{0.5\baselineskip}\centering%
\ifCLASSOPTIONtechnote
   {\bfseries\large\@IEEEcompsoconly{\sffamily}\@title\par}\vskip 1.3em{\lineskip .5em\@IEEEcompsoconly{\sffamily}\@author
   \@IEEEspecialpapernotice\par{\@IEEEcompsoconly{\vskip 1.5em\relax
   \@IEEEtitleabstractindextextbox{\@IEEEtitleabstractindextext}\par
   \hfill\@IEEEcompsocdiamondline\hfill\hbox{}\par}}}\relax
\else
   \vskip0.2em{\EuMWtitlesize\ifCLASSOPTIONtransmag\bfseries\LARGE\fi\@IEEEcompsoconly{\sffamily}\@IEEEcompsocconfonly{\normalfont\normalsize\vskip 2\@IEEEnormalsizeunitybaselineskip
   \bfseries\Large}\@title\par}\vskip1.0em\par
   \ifCLASSOPTIONconference%
      {\@IEEEspecialpapernotice\mbox{}\vskip\@IEEEauthorblockconfadjspace%
       \mbox{}\hfill\begin{@IEEEauthorhalign}\@author\end{@IEEEauthorhalign}\hfill\mbox{}\par}\relax
   \else
      \ifCLASSOPTIONpeerreviewca
         {\@IEEEcompsoconly{\sffamily}\@IEEEspecialpapernotice\mbox{}\vskip\@IEEEauthorblockconfadjspace%
          \mbox{}\hfill\begin{@IEEEauthorhalign}\@author\end{@IEEEauthorhalign}\hfill\mbox{}\par
          {\@IEEEcompsoconly{\vskip 1.5em\relax
           \@IEEEtitleabstractindextextbox{\@IEEEtitleabstractindextext}\par\hfill
           \@IEEEcompsocdiamondline\hfill\hbox{}\par}}}\relax
      \else
         \ifCLASSOPTIONtransmag
           {\@IEEEspecialpapernotice\mbox{}\vskip\@IEEEauthorblockconfadjspace%
            \mbox{}\hfill\begin{@IEEEauthorhalign}\@author\end{@IEEEauthorhalign}\hfill\mbox{}\par
           {\vspace{0.5\baselineskip}\relax\@IEEEtitleabstractindextextbox{\@IEEEtitleabstractindextext}\vspace{-1\baselineskip}\par}}\relax
         \else
           {\lineskip.5em\@IEEEcompsoconly{\sffamily}\sublargesize\@author\@IEEEspecialpapernotice\par
           {\@IEEEcompsoconly{\vskip 1.5em\relax
            \@IEEEtitleabstractindextextbox{\@IEEEtitleabstractindextext}\par\hfill
            \@IEEEcompsocdiamondline\hfill\hbox{}\par}}}\relax
         \fi
      \fi
   \fi
\fi\par\addvspace{0.0\baselineskip}\egroup}

\def\EuMWtitlesize{\@setfontsize{\EuMWtitlesize}{24}{24pt}}
\def\EuMWauthorsize{\@setfontsize{\EuMWauthorsize}{11}{11pt}}
\def\EuMWaffilsize{\@setfontsize{\EuMWaffilsize}{10}{10pt}}
\def\EuMWcaptionsize{\@setfontsize{\EuMWcaptionsize}{9}{10pt}}
\def\EuMWbibsize{\@setfontsize{\EuMWbibsize}{8}{10pt}}

\def\@IEEEauthorblockNstyle{\EuMWauthorsize\@IEEEcompsocnotconfonly{\sffamily}\@IEEEcompsocconfonly{\large}}
\def\@IEEEauthorblockAstyle{\EuMWaffilsize\@IEEEcompsocnotconfonly{\sffamily}\@IEEEcompsocconfonly{\itshape}\@IEEEcompsocconfonly{\large}}
\def\@IEEEauthordefaulttextstyle{\EuMWauthorsize\@IEEEcompsocnotconfonly{\sffamily}\sublargesize}

\def\thebibliography#1{\section*{\refname}%
    \addcontentsline{toc}{section}{\refname}%
    \EuMWbibsize\@IEEEcompsocconfonly{\small}\vskip 0.3\baselineskip plus 0.1\baselineskip minus 0.1\baselineskip
    \list{\@biblabel{\@arabic\c@enumiv}}%
    {\settowidth\labelwidth{\@biblabel{#1}}%
    \leftmargin\labelwidth
    \advance\leftmargin\labelsep\relax
    \itemsep \IEEEbibitemsep\relax
    \usecounter{enumiv}%
    \let\p@enumiv\@empty
    \renewcommand\theenumiv{\@arabic\c@enumiv}}%
    \let\@IEEElatexbibitem\bibitem%
    \def\bibitem{\@IEEEbibitemprefix\@IEEElatexbibitem}%
\def\newblock{\hskip .11em plus .33em minus .07em}%
\ifCLASSOPTIONtechnote\sloppy\clubpenalty4000\widowpenalty4000\interlinepenalty100%
\else\sloppy\clubpenalty4000\widowpenalty4000\interlinepenalty500\fi%
    \sfcode`\.=1000\relax}

\long\def\@makecaption#1#2{%
\ifx\@captype\@IEEEtablestring%
\par\@IEEEtabletopskipstrut
\else
\@IEEEfigurecaptionsepspace
\fi
\setbox\@tempboxa\hbox{\normalfont\footnotesize {#1.}\nobreakspace\nobreakspace #2}%
\ifdim \wd\@tempboxa >\hsize%
\setbox\@tempboxa\hbox{\normalfont\footnotesize {#1.}\nobreakspace\nobreakspace}%
\parbox[t]{\hsize}{\normalfont\footnotesize\noindent\unhbox\@tempboxa#2}%
\else
\ifCLASSOPTIONconference \hbox to\hsize{\normalfont\footnotesize\hfil\box\@tempboxa\hfil}%
\else \hbox to\hsize{\normalfont\footnotesize\box\@tempboxa\hfil}%
\fi\fi
\ifx\@captype\@IEEEtablestring%
\@IEEEtablecaptionsepspace
\else
\fi}

\newlength\tablecaptiontotableskip
\newlength\figuretocaptionskip
\def\@IEEEfigurecaptionsepspace{\vskip\figuretocaptionskip\relax}%
\def\@IEEEtablecaptionsepspace{\vskip\tablecaptiontotableskip\relax}%

\def\abstract{\normalfont%
\@IEEEabskeysecsize\bfseries\textit{\abstractname}\,\bfseries\textit{---}\,%
\@IEEEgobbleleadPARNLSP}%

\def\IEEEkeywords{\normalfont%
\@IEEEabskeysecsize\bfseries\textit{\IEEEkeywordsname}\,\bfseries\textit{---}\,%
\@IEEEgobbleleadPARNLSP}%
\def\endIEEEkeywords{\relax\vspace{0.67ex}%
\par\if@twocolumn\else\endquotation\fi%
\normalsize\normalfont}%

\def\@IEEEauthorblockNtopspace{0ex}
\def\@IEEEauthorblockAtopspace{1mm}
\def\IEEEkeywordsname{Keywords}
\def\subsubsection{\@startsection{subsubsection}{3}{\z@}{1.5ex plus 1.5ex minus 0.5ex}%
{0.7ex plus .5ex minus 0ex}{\normalfont\normalsize\itshape}}%
\newlength{\CPheadmatchindent}%
\def\@seccntformat#1{\hbox to\CPheadmatchindent{\csname the#1dis\endcsname}\hskip 0.1em \relax}
\IEEEilabelindentA \parindent
\IEEEilabelindent \IEEEilabelindentA
\IEEEelabelindent \parindent
\IEEEdlabelindent \parindent
\IEEElabelindent \parindent
\makeatother

    \def\BibTeX{{\rm B\kern-.05em{\sc i\kern-.025em b}\kern-.08em T\kern-.1667em\lower.7ex\hbox{E}\kern-.125emX}}
\begin{document}
 \pagestyle{plain}

	\addtolength{\oddsidemargin}{0.08in}
	\addtolength{\evensidemargin}{0.08in}

	\addtolength{\topmargin}{0.08in}
	\raggedbottom
	\title{Counterfeit Chip Detection using Scattering Parameter Analysis}
	\author{
		Maryam Saadat Safa,
        Tahoura Mosavirik, 
		and Shahin Tajik\\
		Department of Electrical and Computer Engineering,\\ 
        Worcester Polytechnic Institute, Worcester, MA, USA\\
		\{msafa, tmosavirik, stajik\}@wpi.edu

	}
	\maketitle
	%
	%

    \begin{abstract}
The increase in the number of counterfeit and recycled microelectronic chips in recent years has created significant security and safety concerns in various applications. 
Hence, detecting such counterfeit chips in electronic systems is critical before deployment in the field.
Unfortunately, the conventional verification tools using physical inspection and side-channel methods are costly, unscalable, error-prone, and often incompatible with legacy systems.
This paper introduces a generic non-invasive and low-cost counterfeit chip detection based on characterizing the impedance of the system's power delivery network (PDN).
Our method relies on the fact that the impedance of the counterfeit and recycled chips differs from the genuine ones.
To sense such impedance variations confidently, we deploy scattering parameters, frequently used for impedance characterization of RF/microwave circuits.
Our proposed approach can directly be applied to soldered chips on the system's PCB and does not require any modifications on the legacy systems.
To validate our claims, we perform extensive measurements on genuine and aged samples from two families of STMicroelectronics chips to assess the effectiveness of the proposed approach.
  
    \end{abstract}
    

\maketitle

\section{Introduction}
The globalized semiconductor supply chain, developed over the last three decades, has enabled innovation and kept costs low for consumers. 
However, there are single points of failure that could disrupt the supply chain~\cite{SIA_supply_chain}. 
Such disruptions lead to chip shortage for different applications, and consequently, create a surplus of fraudsters and fake parts in the market.
The introduction of counterfeit, recycled, or aged components into the supply chain could lead to quality and performance degradation of electronic systems~\cite{bogus,chopshop}.
F-Secure reported a real-world example of such incidents~\cite{fake_cisco}, where counterfeit Cisco routers have been added to the market and sold to customers.
Counterfeit Xilinx FPGAs, which had ended up on a module used by Boeing for a new U.S. Navy reconnaissance aircraft, is another example of such incidents~\cite{chopshop}.

    
Several techniques have been proposed in the literature to detect counterfeit and recycled chips in a system.
We can categorize these techniques into two classes, namely package/die inspection and behavior analysis.
Infrared (IR) imaging inspection~\cite{ghosh2020automated,dhanuskodi2020counterfoil}, X-ray tomography of ~\cite{shahbazmohamadi2014advanced}, and microwave reflectometry~\cite{Zoughi} are examples of package/die inspections.
Unfortunately, there are several drawbacks and challenges with these verification techniques.
The main shortcoming of inspection techniques is that they cannot verify the electrical functionality of the chip under test and, hence, might leave the counterfeit chips undetected.
Another disadvantage of these inspection methodologies is the lack of scalability due to their customized setup.
        
  \begin{figure*}[t]
     	\centering \noindent
     	\includegraphics[width=18cm]{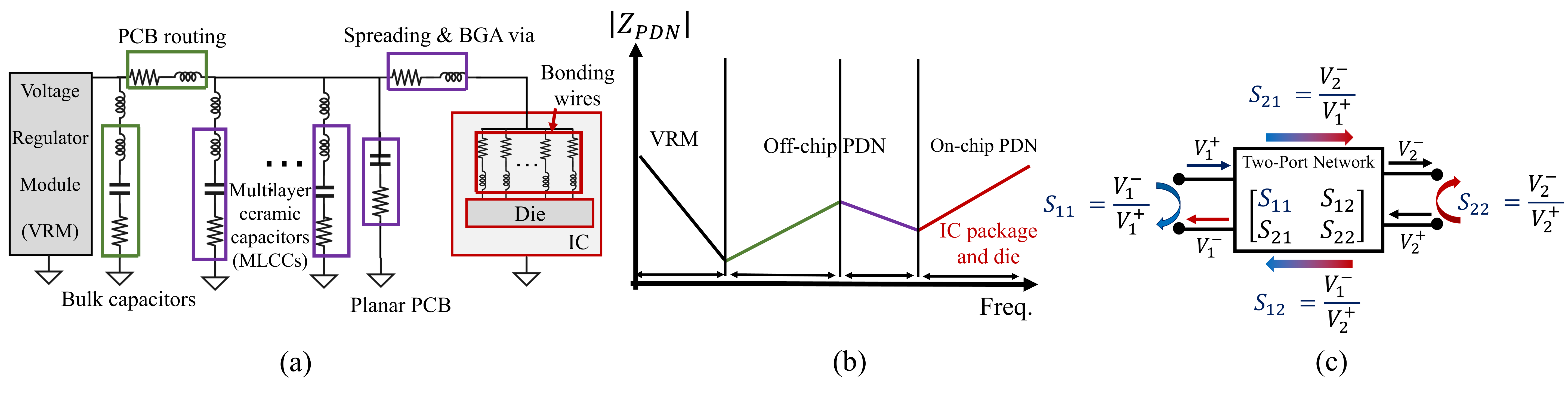}
     	\caption{(a) The equivalent circuit of the PDN of an electronic board~\cite{ImpedanceVerif}. (b) The amplitude of the impedance profile of an electronic board over frequency. (c) Scattering parameters in a two-port network.}
     	\label{Fig_PDN_freq_s_param}
  \end{figure*}

On the other hand, behavior analysis of counterfeit and recycled chips using side-channel analysis (SCA) methods is easier to perform in practice.
The primary assumption in SCA methods is that the counterfeit and recycled chips reveal distinguished behavior in terms of power consumption, EM emanations, and timing, and hence, performing SCA should detect such variations. 
However, the proposed SCA methods for IC counterfeit detection (e.g., ~\cite{Forte,stern2019emforced}) require the execution of test software or activation of pre-designed test circuits to generate traces.
Such requirements might not be compatible with legacy systems.
Besides, SCA methods usually deliver inaccurate and noisy traces for comparisons.
As a result, advanced signal processing and machine learning tools are needed to confidently detect counterfeit chips.

To mitigate the shortcomings of SCA techniques, the verifier can rely directly on measuring the root cause of such side-channel signal variations, i.e., the electrical impedance, rather than performing firmware-dependent side-channel measurements.
The research question here would be how to sense changes in the impedance of a single chip's die on a populated system's board in a non-invasive fashion without a customized and expensive setup. 

    
\noindent{\bfseries Our Contribution:} 
This paper presents a non-invasive and low-cost counterfeit chip detection using scattering parameters, which are standard RF and microwave tools for impedance characterization. 
Our approach is based on the fact that the impedance of the chip's die differs in recycled and counterfeit samples compared to the genuine ones.
In our proposed method, we inject sine wave signals into the power distribution network (PDN) of the system and measure their reflection, which reveals the changes in the impedance.
The impedance of the system's PDN over frequency is affected by various parts of the system~\cite{ScatterVerif,ImpedanceVerif}, from PCB to chip level.
Therefore, by finding the relevant frequency bands, \emph{we can actively probe the impedance of the die in a non-invasive manner.}
We remove the need for using a customized setup and demonstrate how a common portable Vector Network Analyzer (VNA) can precisely detect counterfeit chips without requiring any operator expertise and expensive equipment.
Furthermore, our measurement setup does not require any protected environment or complex system to capture the reflection profile, thus reducing the cost and complexity of the measurements.
To validate our claims, we perform aging on several ST microcontrollers to emulate counterfeit and recycled chips and show that our proposed method can detect these samples with high confidence.
  
 
\section{Background}\label{background}

\subsection{Power Delivery Network (PDN)}\label{sec:pdn}
The power delivery network (PDN) of an electronic system consists of components and interconnects from the voltage regulator module (VRM) to the power rails on the chip.
Fig.~\ref{Fig_PDN_freq_s_param}~(a) demonstrates the equivalent RLC circuit of the PDN model in a typical electronic board~\cite{PDN_2}. 
The PDN functions as a connection between VRM and load circuits of the IC and includes both off-chip (e.g., bulk capacitors, PCB routing, multilayered ceramic capacitors (MLCCs), spreading, and BGA vias) and on-chip (e.g., package and die materials, bonding wires) components~\cite{PDN_2}.
According to Fig.~\ref{Fig_PDN_freq_s_param}~(b), the impedance profile of the PDN has a frequency-dependent behavior. Each component of the PCB has a specific-frequency contribution to the overall impedance profile of the PDN.
In~\cite{ScatterVerif,ImpedanceVerif}, it was demonstrated that the characterization of PDN's impedance in the frequency domain enables the detection of PCB-level tamper events.
Naturally, it is conceivable that any changes in the impedance of the die should also have an impact on the PDN's impedance.



\subsection{Scattering Parameters}\label{sec:scattering}
To characterize the impedance of the PDN, scattering (S) parameters are deployed.
A complex electronic board can be modeled as a one or multiple-port network, (e.g., see in Fig.~\ref{Fig_PDN_freq_s_param}~(c)). 
S-parameters are spectrally measured over the frequency domain to obtain the reflection/transmission properties of the circuit to the applied sine wave voltages/currents.
In Fig.~\ref{Fig_PDN_freq_s_param}~(c), ${V_i^+}$ and ${V_i^-}$ ($i=1,2$) are forward and backward voltage waves through/from the circuit, respectively. 
In frequency domain analysis, these sine waves are represented by frequency, amplitude, and phase.
In our case, we leverage the amplitude response in the frequency domain to accurately sense changes in the impedance of the chip.
To measure the transmitted and/or reflected power of a signal going into and coming back from the PDN at different frequency points, one can employ a VNA.
The impedance profile of the chip can be easily derived from the reflected signal from the PDN.
Eq.~\ref{Conversion_to_Z} represents the relationship between the device under test (DUT) impedance $Z_{DUT}$ and the reflection coefficient $S_{11}$:

\begin{equation}\label{Conversion_to_Z}
Z_{DUT}=Z_0\dfrac{1+S_{11}}{1-S_{11}},
\end{equation}

where $Z_{0}$ is the characteristic impedance of the connecting cables to the VNA.
We only use $S_{11}$ in our proposed approach as the VNA can directly measure it.
However, based on Eq.~\ref{Conversion_to_Z}, it is clear that the reflection coefficient is another representation of the impedance.


    
    

\subsection{Impact of Aging on the Impedance}\label{sec:aging_impact}
\begin{figure}[t!]
     	\centering \noindent
     	\includegraphics[width=0.9\linewidth]{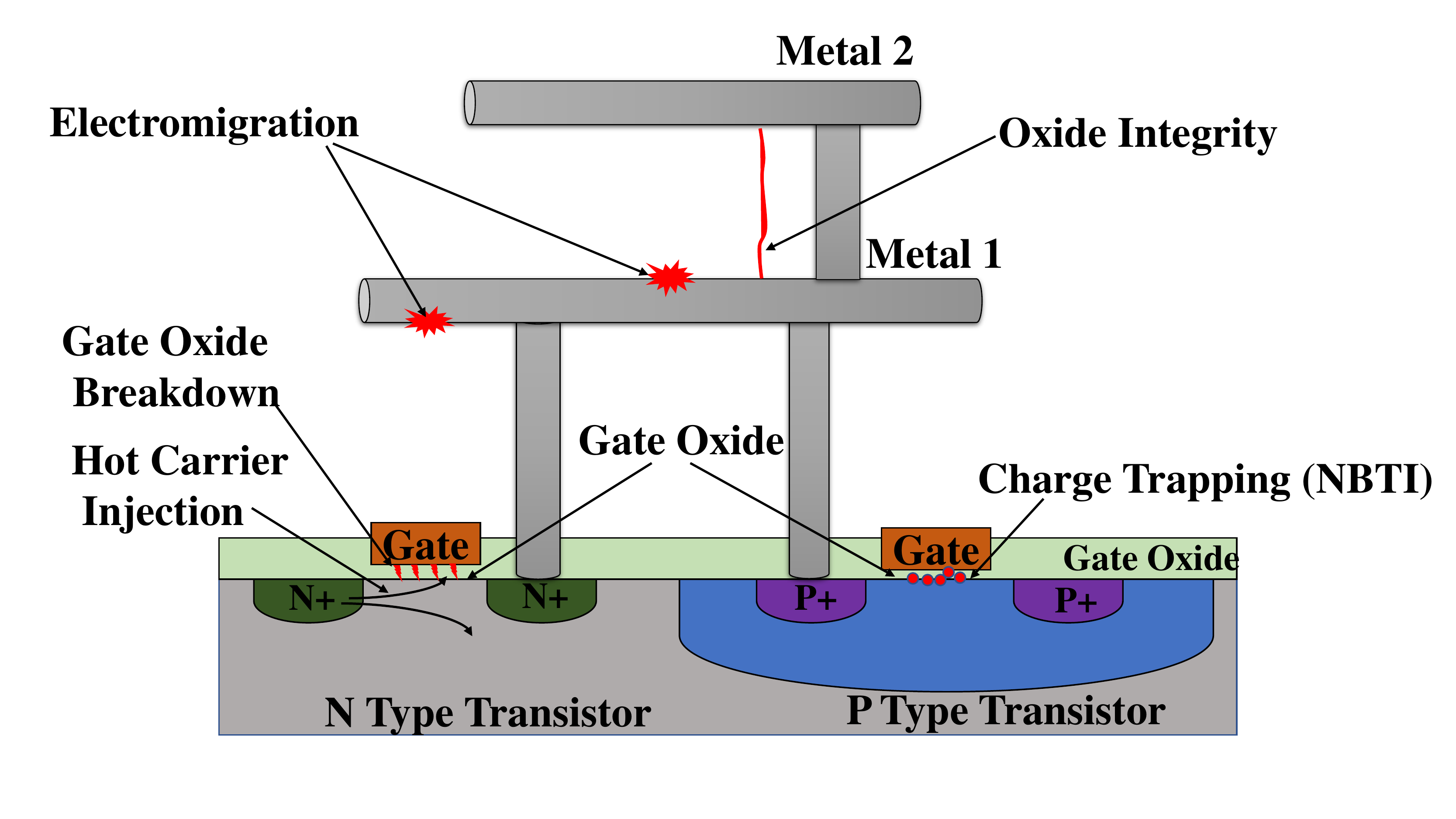}
      \vspace*{-5mm}
     	\caption{Cross-section view of a MOSFET with different aging mechanisms~\cite{boyer}.}
     	\label{Fig_1}
        \end{figure}

In this subsection, we elaborate on how the aging process impacts the transistors' physical parameters, and consequently, the chip's impedance.
Supply voltage, temperature, stress time, and workload are the key contributors to the aging process leading to the device's performance degradation.
The effects causing parameter drifts are negative bias temperature instability (NBTI), hot-carrier injection (HCI), time-dependent dielectric breakdown (TDDB), and electromigration (EM)~\cite{dominic}, see Fig.~\ref{Fig_1}.
Among these parameters, NBTI and HCI are the dominant factors~\cite{dominic}.
HCI is the process that the carriers gain high kinetic energy by an electric field to overcome the potential barrier of the $Si/SiO_2$ interface and leave the channel.
Those carriers could damage the gate oxide layer and get trapped in the oxide shown in Fig.~\ref{Fig_1}~\cite{dominic}.
NBTI, on the other hand, originates from broken Si-H bonds at the interface between the substrate and the gate oxide. 
The NBTI effect increases effective series resistance (ESR) and MOSFETs channel "ON resistance" and decreases capacitance~\cite{NBTI2}.
As a result, the recycled devices show a deviated impedance compared to genuine samples.

\section{Methodology}
\subsection{Threat Model}\label{sec:Methodology_general}   
We make the following assumptions in our threat model.
We assume that the adversary has replaced a genuine chip on a PCB with a counterfeit or recycled one.
The goal of the attacker could be illegal profit or weakening the reliability and lifetime of the system.
We assume that the verifier has access to golden samples for obtaining golden S-parameter signatures.
No control over parts of the design or specific internal support test circuitry for verification is required. 
Finally, we assume that the verifier can stop the clock signal and halt the chip in a specific state for frequency response measurements.
Otherwise, the impedance of the chip could fluctuate during operation~\cite{awal2023utilization}.

\subsection{Scattering Parameters as Signatures}\label{methodology}
As mentioned in Sect.~\ref{sec:aging_impact}, the recycled and counterfeit chips have a deviated impedance compared to genuine samples, and this change would result in the scattering parameters variation of the PDN.
The off-chip components usually affect the PDN impedance up to a few tens of MHz, and the impedance profile at higher frequencies is dominated by the on-chip PDN~\cite{Zhao_Frequency-Domain} (see Fig.~\ref{Fig_PDN_freq_s_param} (b)). 
Consequently, the impact of counterfeit chips on the frequency should be observable at higher frequencies.

\begin{figure}[t!]
     	\centering \noindent
     	\includegraphics[width=7.2cm]{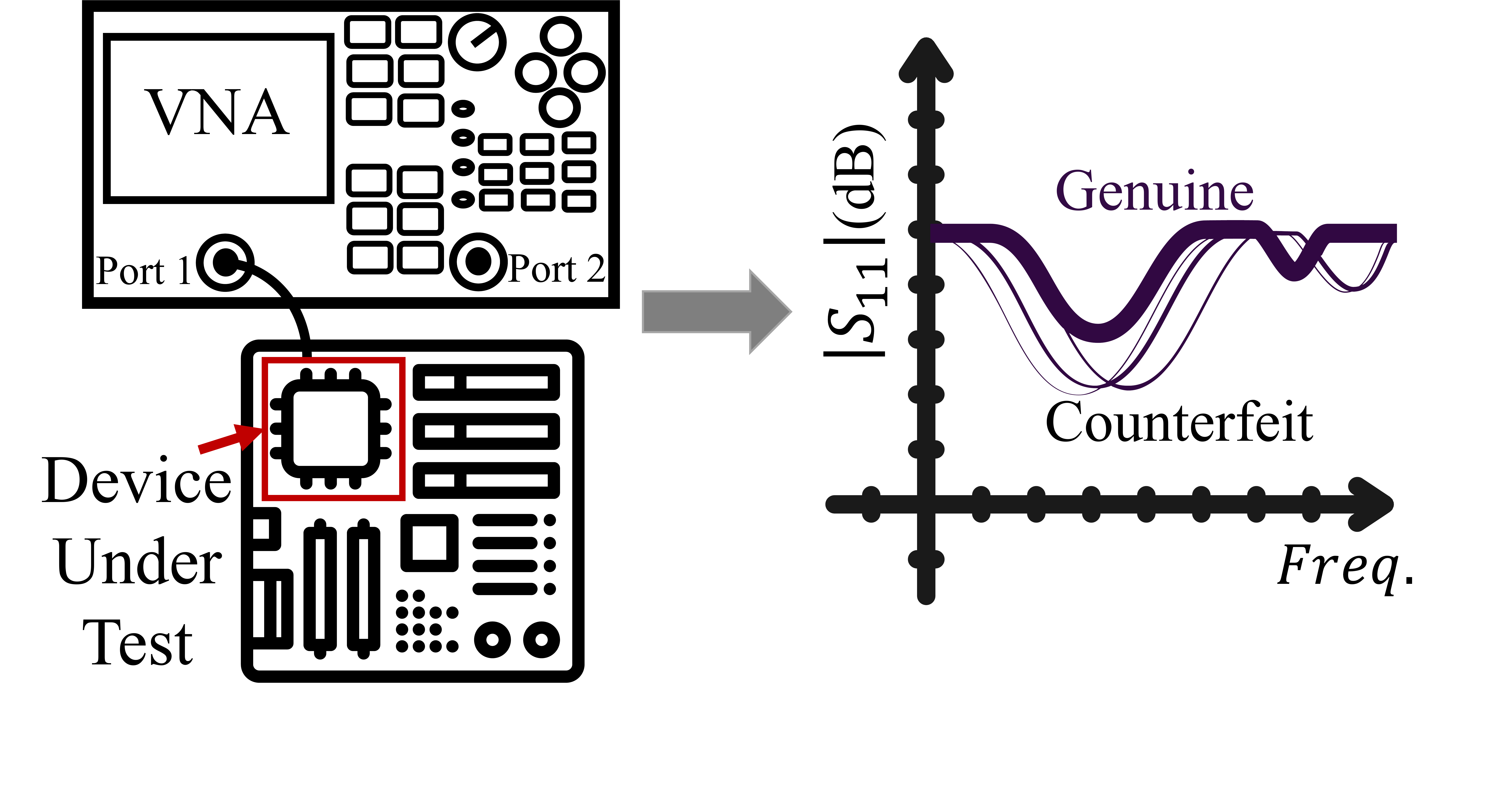}
      \vspace*{-5mm}
     	\caption{Reflection-based hardware signature extraction.}
     	\label{Fig_4}
        \end{figure}

To sense the changes in the PDN impedance, we deploy only the backward-scattered response $S_{11}$ (See Sect.~\ref{sec:scattering}).
Scattering parameters are complex values containing both amplitude and phase; however, to reduce the measurement complexity, we only utilize the amplitude of the reflection scattering parameter ($|S_{11}|={\abs{V_1^-/V_1^+}}$) of the PDN to be able to conduct single-port reflection measurements, practically.
Fig.~\ref{Fig_4} shows the reflection-based hardware signature extraction method proposed here.

As discussed in the previous subsection, we assume that the verifier has access to golden samples to obtain the golden signature.
One of the main challenges in obtaining the golden signature is the noise caused by the environment and measurement uncertainties, as well as the manufacturing process variation between genuine samples.
To mitigate the first type of noise, we needed to integrate our measurement several times at different ambient noise levels and take an average of the responses to increase the signal-to-noise ratio (SNR) and detection confidence. 
Process variation, on the other hand, is addressed by taking an average of the $|S_{11}|$ signatures of all genuine samples at each frequency point and considering it as the reference signature.

To verify the genuinity of a chip, we are taking the following steps.
First, we measure the scattering parameter ($|S_{11}|$) for all genuine chips and take the average of $|S_{11}|$ in three trials to ensure measurement repeatability and reduce any differences caused by environmental noise.
This mean value $\mu$ provides us with the golden signature.
We also calculate the standard deviation $\sigma$ of the measurements for different genuine samples to capture the impact of process variation.
Second, we measure $|S_{11}|$ of the suspicious chip and then compare it with the mean and response of the golden samples.
If the $|S_{11}|$ deviates from the mean, and its amplitude shows a higher deviation than $6\sigma$, we consider the sample not genuine.

\section{Experimental Setup}
\begin{figure}[t!]
     	\centering \noindent
     	\includegraphics[width=0.75\linewidth]{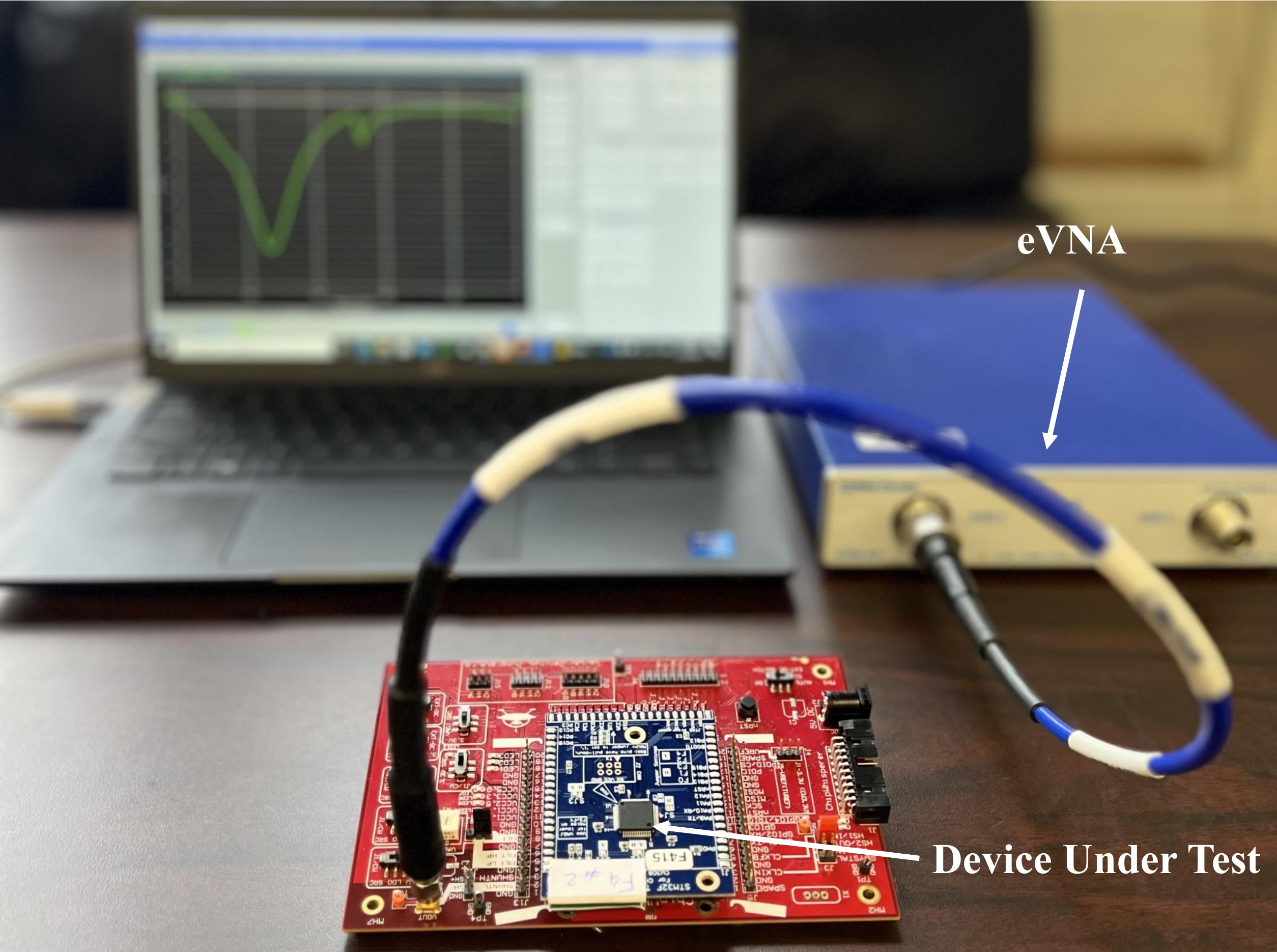}
     	\caption{The experimental setup for reflection response measurement.}
     	\label{fig:experimental_setup}
        \end{figure}

\subsection{Device under Test (DUT)}
For our experiments, we chose the NewAE CW308 UFO boards~\cite{CW308} consisting of base and target boards.
The direct access to the PDN on CW308 boards through an SMA connector was the primary reason for the selection of these kits.
The target boards are detachable and can be easily mounted on the baseboards.
This feature makes it convenient to mount different target boards from the same family on the baseboard and conduct different experiments.
We utilized the STM32F target boards~\cite{STM32F} that include STMicroelectronics STM32F series of ARM Cortex devices for our experiments.
More precisely, we used NAE-CW308T-STM32F3 (12 samples) and 
NAE-CW308T-STM32F4HWC (10 samples) target boards containing STM32F303RCT7 and STM32F415RGT6 microcontrollers.
The nominal voltage of these chips is 3.3 V.
In all accelerated aging experiments, we used the internal clock of the microcontrollers.

\subsection{Measurement Setup}


\begin{figure}[t!]
     	\centering \noindent
     	\includegraphics[width=7cm]{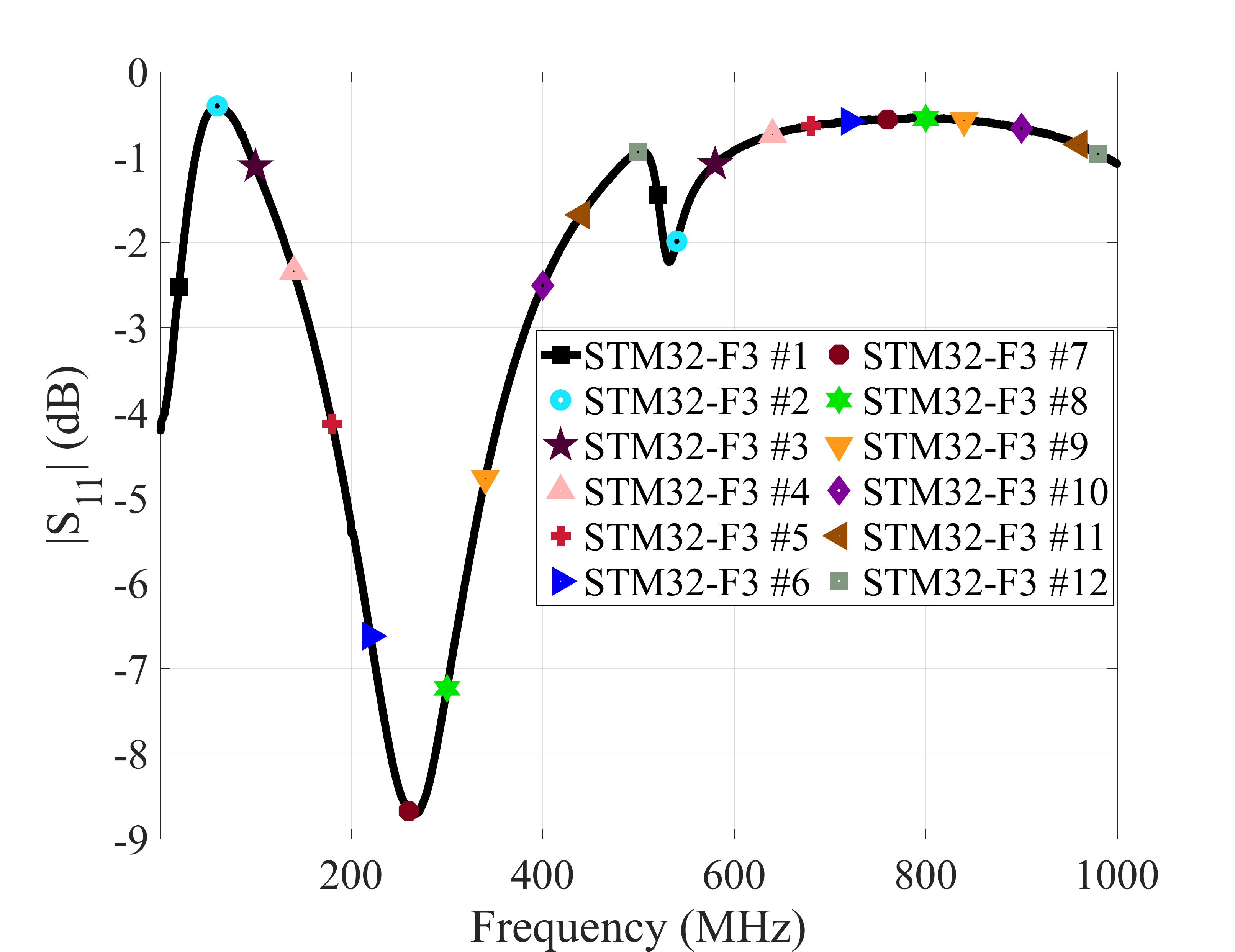}
     	\caption{The amplitude of the reflection response for 12 genuine STM32-F3 samples (powered-off) over 1 MHz - 1 GHz bandwidth.}
     	\label{STM32_F3_all_twelve}
        \end{figure}

\begin{figure}[t!]
     	\centering \noindent
     	\includegraphics[width=7cm]{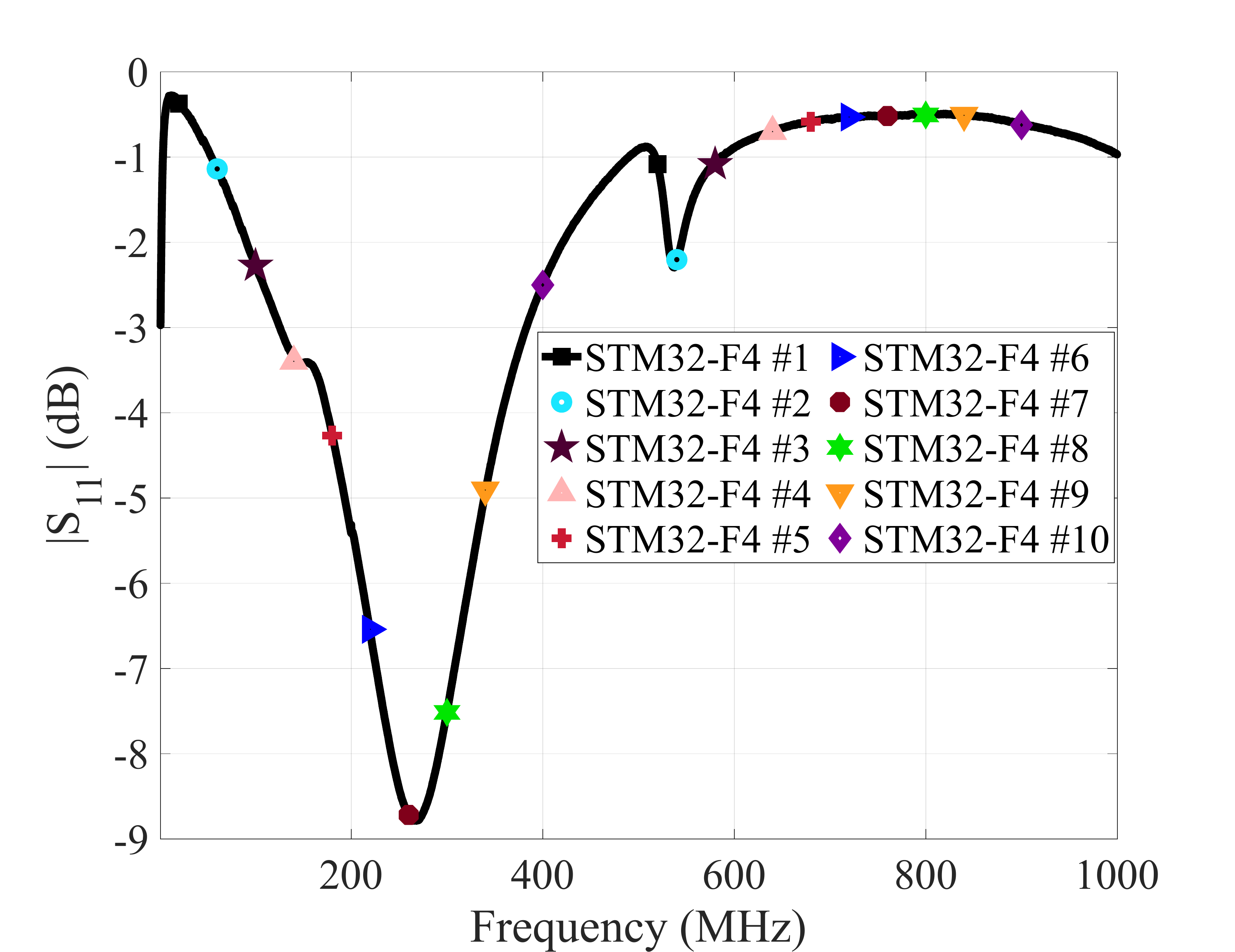}
     	\caption{The amplitude of the reflection response for 10 genuine STM32-F4HWC samples (powered-off) over 1 MHz - 1 GHz bandwidth.}
     	\label{STM32_F4_all_ten}
        \end{figure}

We utilized Mini-circuits eVNA-63+, which is a portable VNA capable of operating within 300 kHz - 6 GHz bandwidth.
The VNA consists of an internal capacitor to filter out the DC voltage, and thus, no external bias tee is required.
We used shielded precision test cables CBL-2FT-SMNM+, which has a male SMA connector on the side of the DUT, and thus, enables the direct connection to the boards without any additional adaptors.
We precisely calibrated the VNA until the SMA connection plane on the baseboards, using open-short-load (OSL) calibration, which is the standard calibration for the one-port reflection/impedance measurements.
We conducted $|S_{11}|$ measurements within 1 MHz - 1 GHz bandwidth.
We set the number of frequency samples to 5000 equally-spaced points to ensure the maximum spectral resolution.
We configured the VNA to carry out the measurements with 10 kHz IF bandwidth and 5 dBm output power level.

\begin{figure}[t!]
     	\centering \noindent
     	\includegraphics[width=7cm]{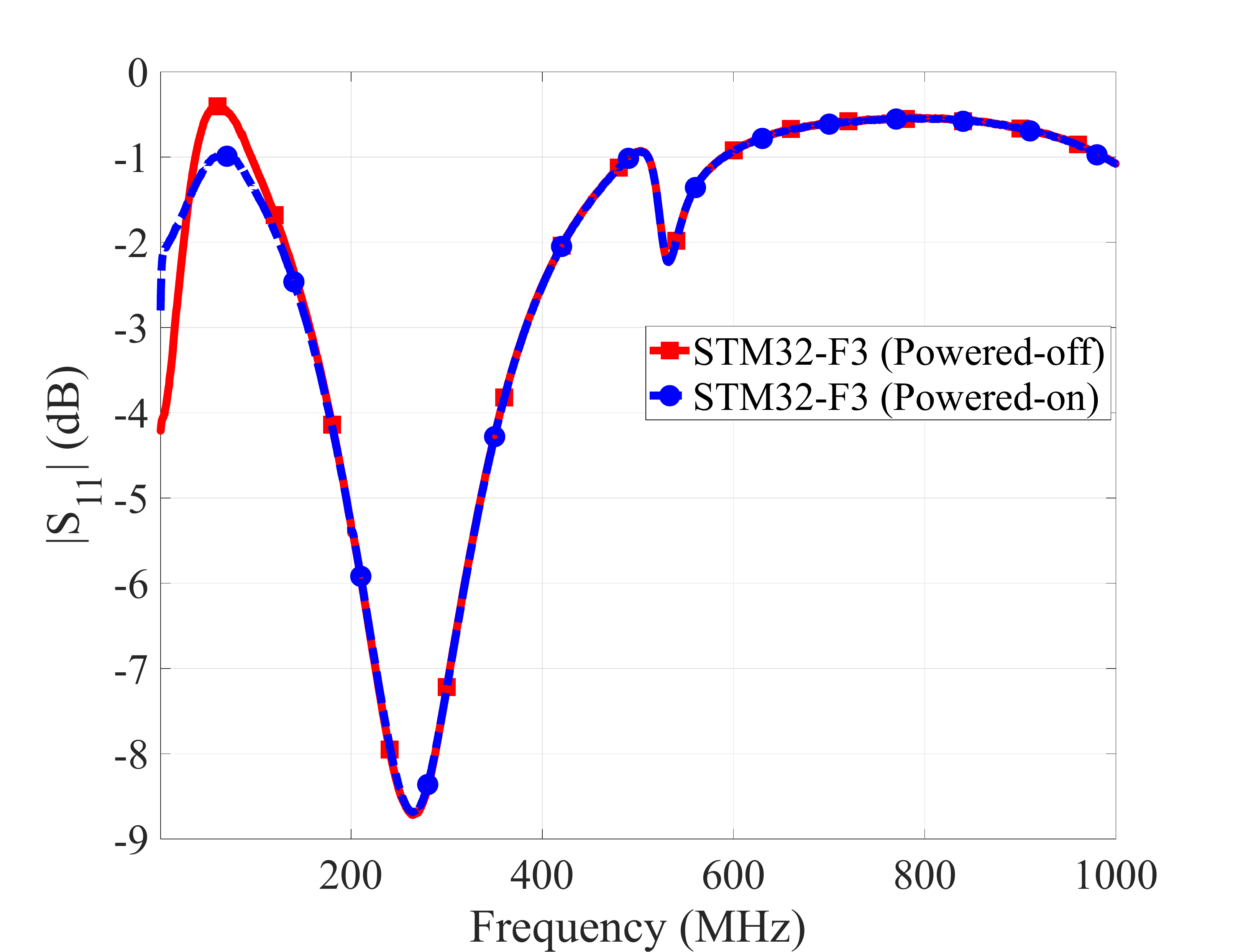}
     	\caption{The reflection profile of the genuine STM32-F3 sample for powered-on and powered-off cases within 1 MHz to 1 GHz band.}
     	\label{STM32_F3_Off_On}
        \end{figure}

\begin{figure}[t!]
     	\centering \noindent
     	\includegraphics[width=7cm]{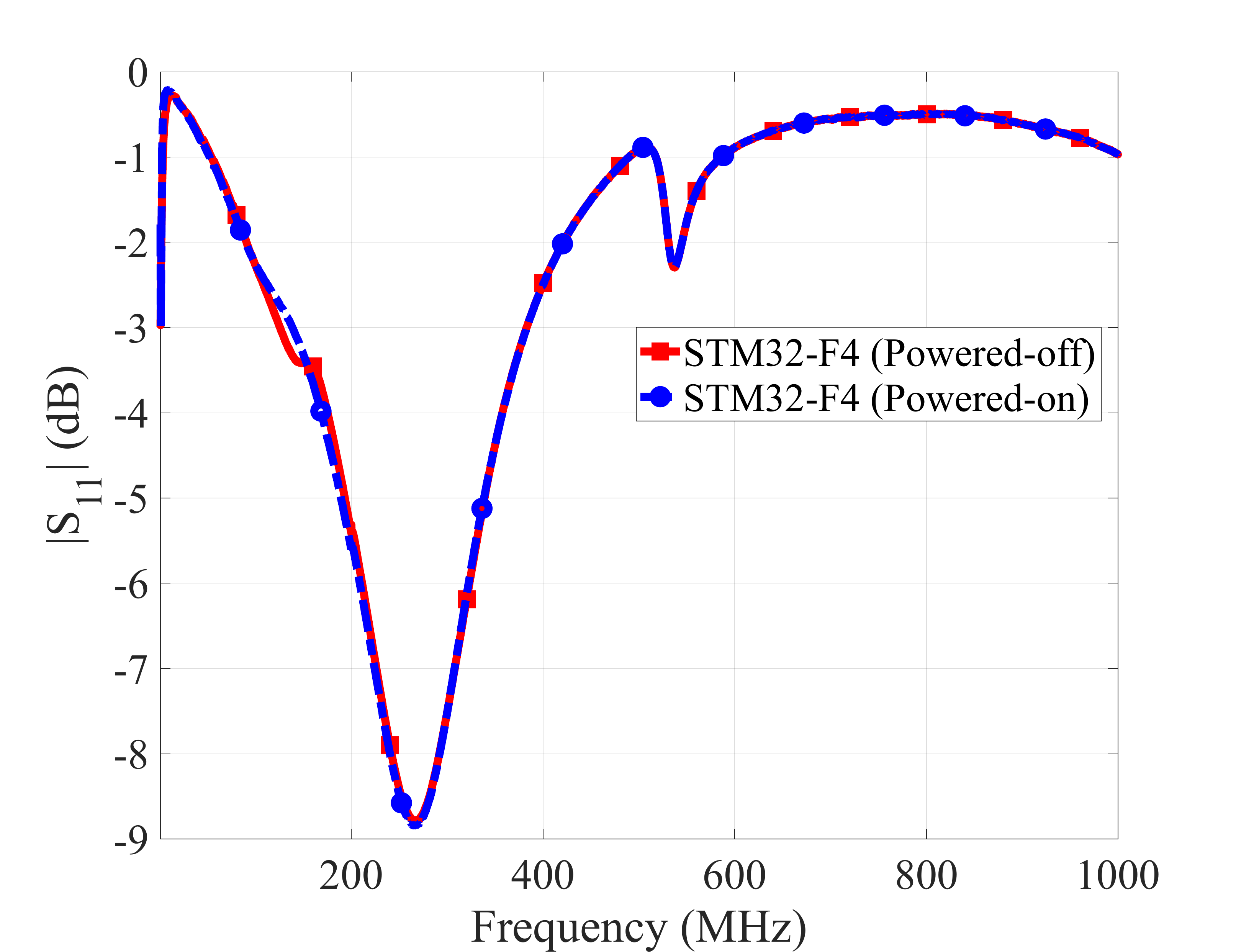}
     	\caption{The reflection profile of the genuine STM32-F4HWC sample for powered-on and powered-off cases within 1 MHz - 1 GHz band.}
     	\label{STM32_F4_Off_On}
        \end{figure}

\section{Results}\label{results}

\subsection{Genuine Chips Characterization}
Our first experiment investigates the signature consistency between identical boards (genuine) from the same family.
To this end, we performed $|S_{11}|$ measurements on the 3.3 V voltage rails of the 12 genuine STM32-F3 and 10 STM32-F4HWC samples.
It should be noted that in all measurements, the chips were powered on, but no instructions were being executed on the chip.
We performed each measurement ten times for each sample and took the average signature of the measured signatures at each frequency sample within 1 MHz - 1 GHz to reduce the environmental variation and noise effects on the experiments.
Fig.~\ref{STM32_F3_all_twelve} and~\ref{STM32_F4_all_ten} show the $|S_{11}|$ signatures of twelve STM32-F3 and ten STM32-F4HWC samples when the chips are powered-off.  
The $|S_{11}|$ results of these genuine samples confirm the signature consistency between them.
The slight difference between these signatures results from the chip's manufacturing process variation.

In the next experiment, we aim at investigating the effect of the chip's die on the impedance profile. 
We performed two $|S_{11}|$ measurements on one of our genuine chips in powered-on and off states. 
Fig.~\ref{STM32_F3_Off_On} and~\ref{STM32_F4_Off_On} depict the reflection profile for the genuine STM32-F3 and STM32-F4HWC samples over the band of 1 MHz - 1 GHz, respectively.
One interesting observation is that in specific frequency bands, such as 50 MHz - 120 MHz for STM32-F3 and 70 MHz - 130 MHz for STM32-F4HWC, $|S_{11}|$ signature has been changed significantly due to the effect of powering up the chip.
The main reason for such a change is the addition of the ON-transistors' impedance to the PDN circuit~\cite{smith2011capacitance}.
Therefore, we expect to observe the impact of aging on the PDN's impedance profile in these frequency bands as well.
      
\begin{figure}[t!]
\centering \noindent
\includegraphics[width=9cm]{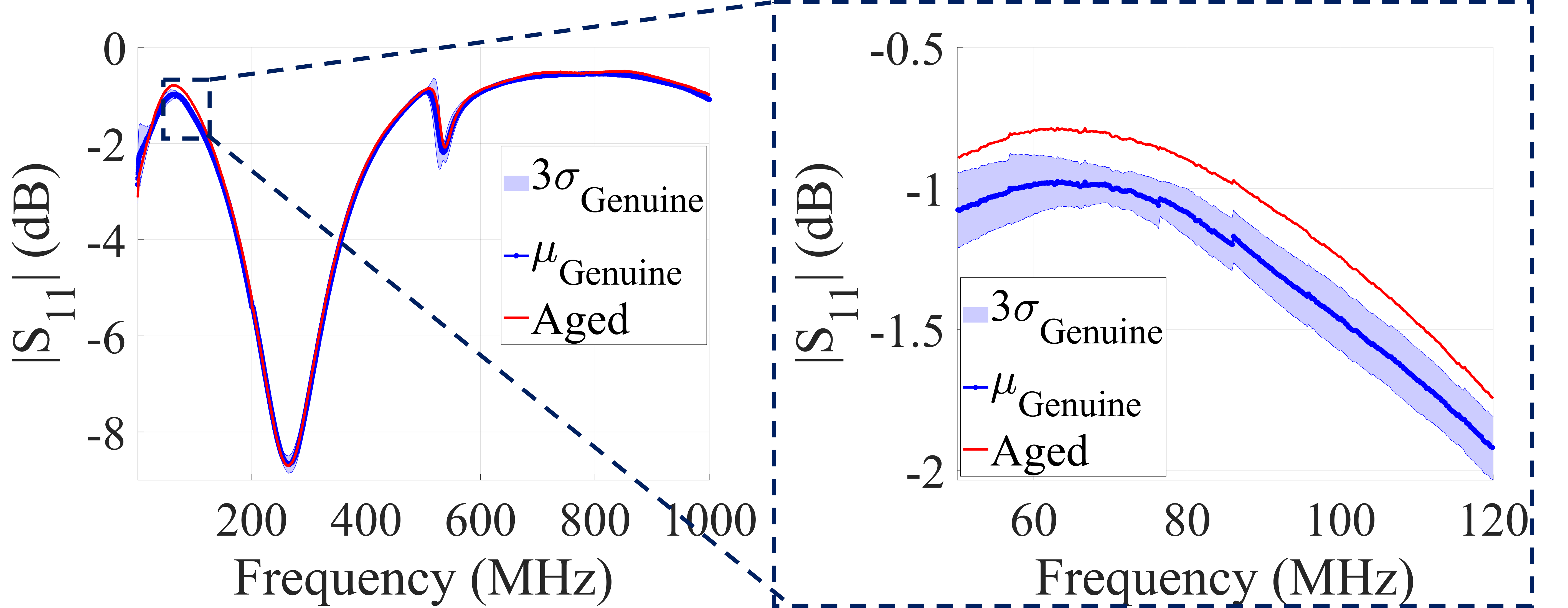}
\caption{The mean and $3\sigma$ (from each side of the mean) of $|S_{11}|$ response for the genuine STM32-F3 samples and $|S_{11}|$ response for the aged chip (powered-on state). The right-side figure shows the zoomed-in view of the bandwidth (60 MHz - 120 MHz) with a high deviation from the mean graph.}
\label{STM32f3_result}
\end{figure}
        
\subsection{Accelerated Aging}
Next, we performed accelerated aging experiments on the target chips to replicate the effect of recycling on the chip.
Accelerated aging includes applying high-level stress conditions, including high-level voltage or temperature (higher than the nominal voltage or temperature), for a short period of time to accelerate the damage due to aging. 
To perform the aging test, the voltage supplied to the STM32F samples was 5.5 V, which is 2.2 V higher than the nominal operating voltage (3.3 V) of the DUTs.
We let the target devices go through the aging process for 216 hours at room temperature.
   
As mentioned in Sect.~\ref{background}, the chip's workload is a crucial parameter in aging. 
Working under a heavy workload accelerates aging deterioration because a larger area of the chip's circuit is involved.
During the aging process, a firmware, containing AES-256 encryption instructions was running on the microcontrollers, and the encrypted output was XORed by four random numbers in four consecutive commands.
Finally, the results of XOR operations were monitored by LEDs on the CW308 baseboards for health monitoring of the chips under aging.
The voltage was reset to 3.3 V for $|S_{11}|$ measurements.
It should be noted that loading a firmware into the microcontrollers could alter the chip's impedance as well~\cite{awal2023utilization}. 
To eliminate this effect, the chip is subjected to the reference firmware after aging so that any difference in impedance or S-parameter is only due to the aging effect. 
To reduce the impact of thermal noise, we recorded all measurements under the nominal operating conditions with an hour thermal stabilization period after the aging test.
   
It is worth mentioning that during the aging test, the baseboards were not aged as we directly supplied the voltage to the target boards.
Therefore, if a change in the S-parameter of the chip is detected, we can confidently attribute this change to the aging effect.
Finally, we measured the aged samples' $|S_{11}|$ signatures to evaluate the aging repercussion on $|S_{11}|$ signatures.

\begin{figure}[t!]
\centering \noindent
\includegraphics[width=9cm]{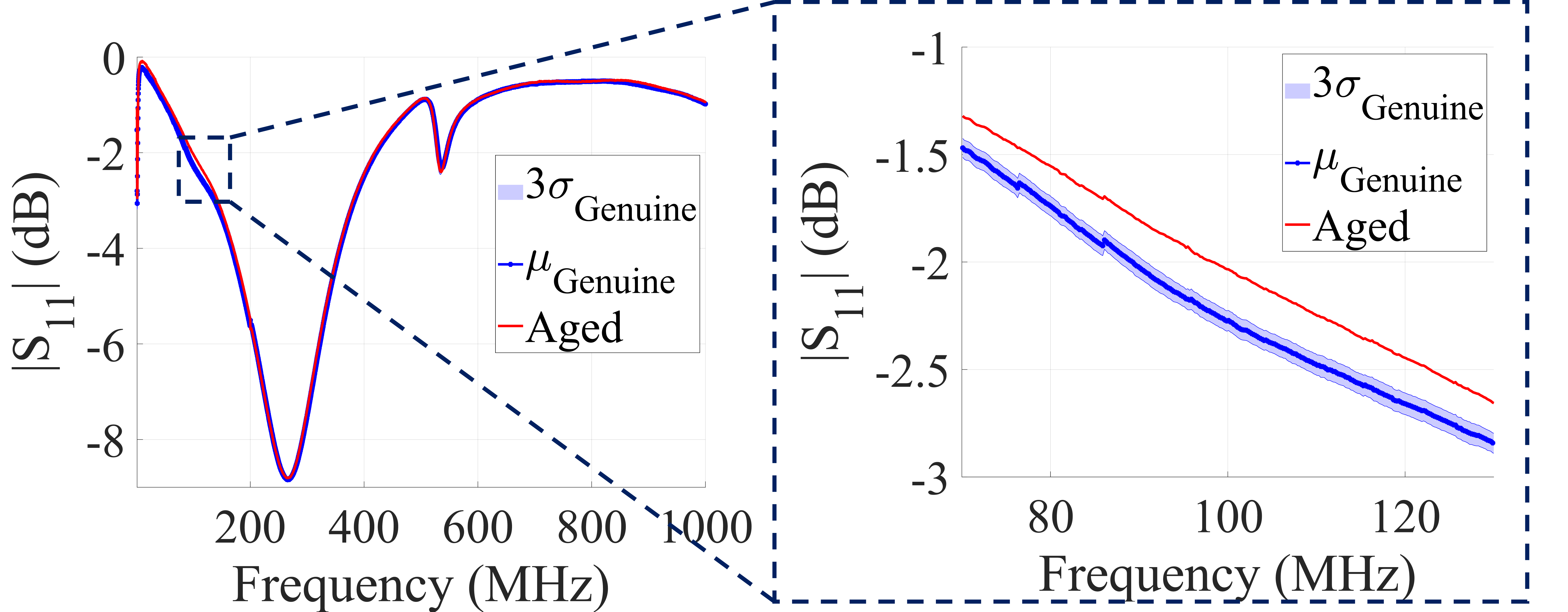}
\caption{The mean and $3\sigma$ of $|S_{11}|$ response for the genuine STM32-F4HWC samples and $|S_{11}|$ response for the aged chip (powered-on state). The right-side figure shows the zoomed-in view of the bandwidth (70 MHz - 130 MHz) with a high deviation from the mean graph.}
\label{STM32f4_result}
\end{figure}

\subsection{Counterfeit Chip Detection}

In this section, we deploy the measured $|S_{11}|$ responses for detecting counterfeit (aged) chips. 
In the first scenario, where we replicated the aging effect on one chip from each STM32F family, we assessed the efficacy of our proposed detection method to detect these aged samples.
We performed $|S_{11}|$ measurement on the powered-on aged samples ten times and took the average of the measurement responses at each frequency sample.
Thereafter, the average $|S_{11}|$ profile of the genuine chips is compared to the signature of the aged chip.
For a better illustration of the differences between genuine and aged chips, the mean and six times standard deviation of $|S_{11}|$ values for genuine STM32-F3 and STM32-F4HWC chips and their corresponding aged chip responses are given in Fig.~\ref{STM32f3_result} and~\ref{STM32f4_result}. 
The shaded area (6$\sigma$) represents the impact of process variation and encloses the area wherein all genuine chip signatures are expected to occur.
According to Fig.~\ref{STM32f3_result} and~\ref{STM32f4_result}, the chip aging has an impact on the portion of the spectrum where we detected the chip's contribution in powered-on and powered-off measurements (cf. Fig.~\ref{STM32_F3_Off_On} and~\ref{STM32_F4_Off_On}).
The zoomed-in view graphs in Fig.~\ref{STM32f3_result} and Fig.~\ref{STM32f4_result} show how much the aged chip signatures deviate from the mean of the genuine chip's signatures. 
Since the aged chips' responses do not overlap with the shaded blue areas in particular frequency bands (e.g., zoomed-in graphs), we can conclude that the deviations do not result from process variation in these bands
This observation confirms that the chip's impedance changes due to the aging effect.
Furthermore, the deviation due to aging increases as the aging process time increases.
      
\begin{figure}[t!]
     	\centering \noindent
     	\includegraphics[width=9cm]{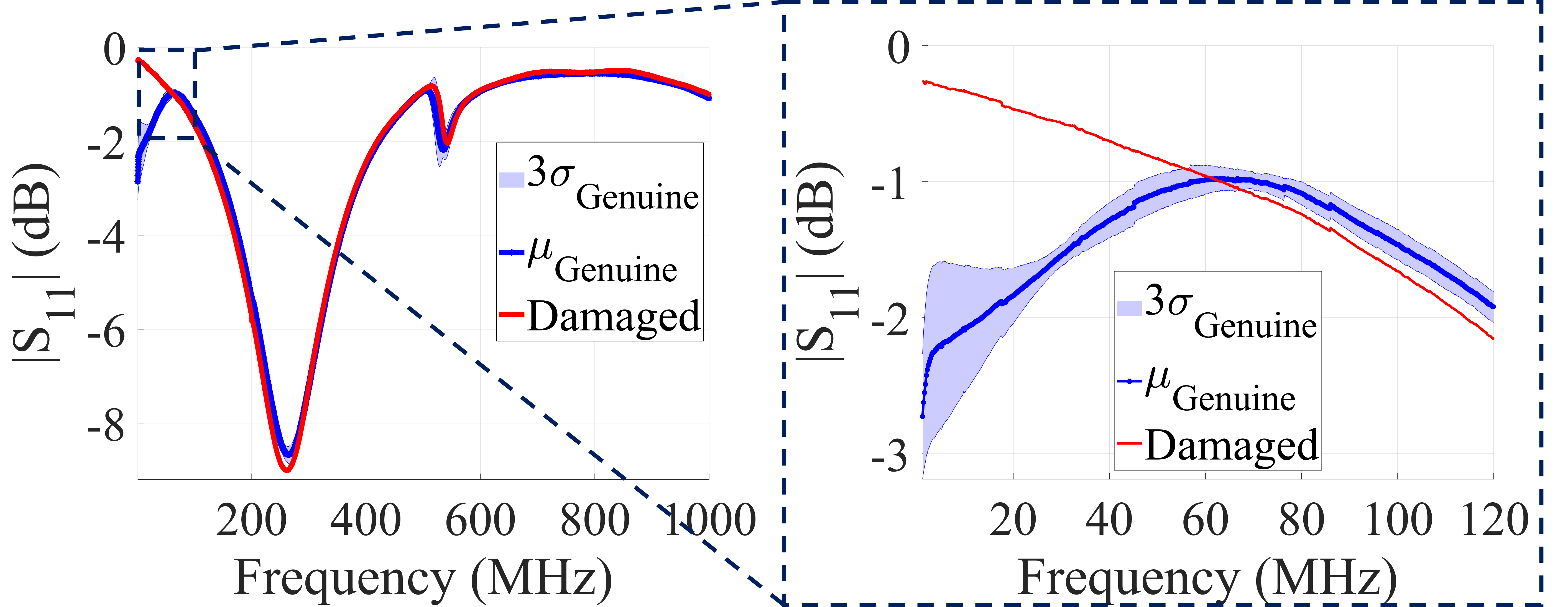}
     	\caption{The mean and $3\sigma$ of $|S_{11}|$ response for the genuine STM32-F3 chips and $|S_{11}|$ response for the damaged chip (powered-on condition). The right-side figure shows the zoomed-in view of the bandwidth (1 MHz - 120 MHz) with a high deviation from the mean graph.}
     	\label{STM32f3_damaged}
        \end{figure}

In the second scenario, to assess the effectiveness of the proposed framework, we deployed the proposed method to characterize a damaged chip from the STM32-F3 family. 
We prepared such a chip by exposing it to a high voltage level (6 V), leading to its breakdown. 
We measured the $|S_{11}|$ parameter of the damaged chip and compared it with the average $|S_{11}|$ signatures of the genuine samples. 
This effect is visible in Fig.~\ref{STM32f3_damaged}, where the damaged chip's signature deviates from the mean of the genuine signatures.
However, this attack affects a larger bandwidth and mostly lower frequencies compared to the aging effect shown in the zoomed-in view of Fig.~\ref{STM32f3_result}. 
A possible explanation could be when the chip is damaged, a larger area of the chip has been exposed to degradation.
The effect of the larger parts is observed in lower frequencies, as the wavelength is inversely proportional to the frequency.      
A slight resonance frequency shift is also observed at a higher frequency, which is due to the change in capacitance and inductance of the PDN, as the resonance frequency is related to the capacitance and inductance of the PDN circuit. 
Overall, our experimental results show the proposed method is capable of confidently detecting counterfeit and recycled samples at distinct bands.

\section{Conclusion}
We presented a non-invasive and low-cost counterfeit chip detection approach based on characterizing the reflection frequency response of the chip's PDN.
Unlike the conventional SCA detection approaches, this method makes it possible to detect counterfeiting without requiring running any test firmware or integrating additional test circuitry onto the chip, which makes it compatible with legacy systems.
We measured the $|S_{11}|$ values of the chips' PDN using a portable VNA.
Afterward, $|S_{11}|$ signatures of aged and genuine DUTs were compared to demonstrate that chip counterfeiting and recycling attacks would impact the reflection profile of the PDN at certain frequency bands.
Based on our acquired results, the counterfeit chips could be detected confidently, without applying any signal processing or machine learning algorithms.
The proposed technique can be considered complementary to the \emph{ScatterVerif} framework, presented in~\cite{ScatterVerif}, enabling the verification of the entire system, from PCB to chip-level, in a unified manner.

\section*{Acknowledgment}
This work was sponsored by Electric Power Research Institute (EPRI).


\bibliographystyle{ieeetr}
\bibliography{references}

\begin{thebibliography}{10}

\bibitem{SIA_supply_chain}
{Semiconductor Industry Association}, ``{Strengthening the Global Semiconductor
  Supply Chain in an Uncertain Era},'' 2021.

\bibitem{bogus}
M.~Pecht and S.~Tiku, ``Bogus: electronic manufacturing and consumers confront
  a rising tide of counterfeit electronics,'' {\em IEEE spectrum}, vol.~43,
  no.~5, pp.~37--46, 2006.

\bibitem{chopshop}
J.~Villasenor and M.~Tehranipoor, ``Chop shop electronics,'' {\em IEEE
  Spectrum}, vol.~50, no.~10, pp.~41--45, 2013.

\bibitem{fake_cisco}
D.~Janushkevich, ``{The Fake Cisco: Hunting for Backdoors in Counterfeit Cisco
  Devices},'' {\em F-Secure Consulting, Hardware Security Team}, 2020.

\bibitem{ghosh2020automated}
P.~Ghosh, U.~J. Botero, F.~Ganji, D.~Woodard, R.~S. Chakraborty, and D.~Forte,
  ``Automated detection and localization of counterfeit chip defects by texture
  analysis in infrared {(IR)} domain,'' in {\em 2020 IEEE Physical Assurance
  and Inspection of Electronics (PAINE)}, pp.~1--6, IEEE, 2020.

\bibitem{dhanuskodi2020counterfoil}
S.~N. Dhanuskodi, X.~Li, and D.~Holcomb, ``$\{$COUNTERFOIL$\}$: Verifying
  provenance of integrated circuits using intrinsic package fingerprints and
  inexpensive cameras,'' in {\em 29th USENIX Security Symposium (USENIX
  Security 20)}, pp.~1255--1272, 2020.

\bibitem{shahbazmohamadi2014advanced}
S.~Shahbazmohamadi, D.~Forte, and M.~Tehranipoor, ``Advanced physical
  inspection methods for counterfeit {IC} detection,'' in {\em ISTFA 2014:
  Conference Proceedings from the 40th International Symposium for Testing and
  Failure Analysis}, p.~55, ASM International, 2014.

\bibitem{Zoughi}
S.~Shinde, S.~Jothibasu, M.~T. Ghasr, and R.~Zoughi, ``Wideband microwave
  reflectometry for rapid detection of dissimilar and aged {IC}s,'' {\em IEEE
  Transactions on Instrumentation and Measurement}, vol.~66, no.~8, 2017.

\bibitem{ImpedanceVerif}
T.~Mosavirik, P.~Schaumont, and S.~Tajik, ``Impedanceverif: On-chip impedance
  sensing for system-level tampering detection,'' {\em IACR Transactions on
  Cryptographic Hardware and Embedded Systems}, vol.~1, pp.~301--325, 2023.

\bibitem{Forte}
M.~M. Alam, M.~Tehranipoor, and D.~Forte, ``{Recycled {FPGA} Detection Using
  Exhaustive {LUT} Path Delay Characterization and Voltage Scaling},'' {\em
  IEEE Transactions on Very Large Scale Integration (VLSI) Systems}, vol.~27,
  no.~12, pp.~2897--2910, 2019.

\bibitem{stern2019emforced}
A.~Stern, U.~Botero, F.~Rahman, D.~Forte, and M.~Tehranipoor, ``{EMFORCED}:
  Em-based fingerprinting framework for remarked and cloned counterfeit {IC}
  detection using machine learning classification,'' {\em IEEE Transactions on
  Very Large Scale Integration (VLSI) Systems}, vol.~28, no.~2, pp.~363--375,
  2019.

\bibitem{ScatterVerif}
T.~Mosavirik, F.~Ganji, P.~Schaumont, and S.~Tajik, ``Scatterverif:
  Verification of electronic boards using reflection response of power
  distribution network,'' {\em ACM Journal on Emerging Technologies in
  Computing Systems}, vol.~18, no.~4, pp.~1--24, 2022.

\bibitem{PDN_2}
B.~Zhao, C.~Huang, K.~Shringarpure, J.~Fan, B.~Archambeault, B.~Achkir,
  S.~Connor, M.~Cracraft, M.~Cocchini, A.~Ruehli, and J.~Drewniak, ``Analytical
  {PDN} voltage ripple calculation using simplified equivalent circuit model of
  {PCB PDN},'' pp.~133--138, 2015.

\bibitem{dominic}
D.~Lorenz, {\em Aging Analysis of Digital Integrated Circuits,}.
\newblock PhD thesis, Technische Universität München, 2012.

\bibitem{NBTI2}
D.~V. M.~A.~Alam, H.~Kufluoglu and S.~Mahapatra., ``A comprehensive model for
  {PMOS NBTI} degradation,'' June 2007.

\bibitem{boyer}
S.~B. D. L.~G. lexandre Boyer, Amadou Cisse~Ndoye and B.~Vrignon,
  ``Characterization of the evolution of {IC} emissions after accelerated
  aging,'' {\em {IEEE} Transactions on Electromagnetic Compatibility}, 2009.

\bibitem{awal2023utilization}
M.~S. Awal, C.~Thompson, and M.~T. Rahman, ``Utilization of impedance disparity
  incurred from switching activities to monitor and characterize firmware
  activities,'' {\em arXiv preprint arXiv:2301.06799}, 2023.

\bibitem{Zhao_Frequency-Domain}
S.~Zhao, I.~Ahmed, V.~Betz, A.~Lotfi, and O.~Trescases, ``Frequency-domain
  power delivery network self-characterization in {FPGAs} for improved system
  reliability,'' {\em {IEEE} Transactions on Industrial Electronics}, vol.~65,
  no.~11, pp.~8915--8924, 2018.

\bibitem{CW308}
{NewAE Technology Inc.}, ``{CW308 UFO Target Board Datasheet}.'' url:
  https://media.newae.com/datasheets/NAE-CW308-datasheet.pdf, 2018.

\bibitem{STM32F}
{NewAE Technology Inc.}, ``{STM32F Boards Datasheet}.'' url:
  https://www.st.com/resource/en/datasheet/stm32f303vc.pdf, 2016.

\bibitem{smith2011capacitance}
L.~Smith, S.~Sun, M.~Sarmiento, Z.~Li, and K.~Chandrasekar, ``On-die
  capacitance measurements in the frequency and time domains,'' {\em Presented
  at the DesignCon 2011}, 2011.

\end{thebibliography}

\end{document}